\documentclass[intlimits,twoside,a4paper]{article}

\usepackage{amsmath,amssymb}
\usepackage{graphicx}
\usepackage{wrapfig}
\usepackage{multirow} 

\usepackage[cp1251]{inputenc}

\usepackage[eqsecnum]{cmpj3}

\issue{2018}{21}{1}{13702}
\doinumber{10.5488/CMP.21.13702}

\title[Performance comparison of 2-1-3, 1-3 and 1-3-2 piezoelectric composite transducers]%
{Performance comparison of 2-1-3, 1-3 and 1-3-2 piezoelectric composite transducers by finite element method\thanks{Supported by ``the Fundamental Research Funds for the Central Universities(BLX2015-24)''.}}
\author[Y. Sun, B. Hua]{Y. Sun\refaddr{label1}\footnote{Corresponding author, Tel: +8615910454437, E-mail: sunyang@bjfu.edu.cn.}\,, B. Hua\refaddr{label2}}
\addresses{
\addr{label1} College of Science, Beijing Forestry University, \\ No.~35 Tsinghua East Road, Haidian District, 100083 Beijing, China
\addr{label2}Systems Engineering Research Institute, China State Shipbuilding Corporation, \\ 1 Fengxian East Road, Haidian District, 100094 Beijing, China}

\date{Received August 17, 2017, in final form December 13, 2017}

\begin{document}

\maketitle

\begin{abstract}
1-3 type, 1-3-2 type and 2-1-3 type piezoelectric composites are three proper smart materials for the design and manufacture of ultrasonic transducers. They have been proposed in different stages but possess similar properties. Compared with the initial 1-3 type composite, 1-3-2 composite is of higher mechanical stability. Compared with 1-3-2 composite, 2-1-3 composite has lower manufacturing difficulty. In this paper, a comparative study on these three composites in terms of receiving transducer material properties is presented. Finite element method (FEM) has been adopted to calculate longitudinal velocity, thickness electromechanical coupling coefficient and voltage receiving sensitivity. It is concluded that for a large aspect ratio $\alpha=1$, the performance of 2-1-3 composite transducer is much better than that of 1-3 and 1-3-2 composite transducers. The thickness electromechanical coupling coefficient of 2-1-3 composite transducer is about 5.58 times that of 1-3 composite transducer and 7.42 times that of 1-3-2 composite transducer. The voltage receiving sensitivity at 2 kHz of 2-1-3 composite transducer is 13.1 dB higher than that of 1-3-2 composite transducer and 12.3 dB higher than that of 1-3 composite transducer.
\pacs 77.84.-s, 43.30.Yj, 02.70.Dh, 77.84.Cg, 81.05.Qk, 77.65.-j
\end{abstract}

\section{Introduction}

In the field of ultrasonic transducers, 1-3 piezoelectric composite has been widely used as a kind of smart material instead of piezoelectric ceramics because of its better flexibility, lower acoustic impedance and higher piezoelectricity \cite{Rit00,Akd05,Ram14,Yan15,Rao10}. Generally, piezoelectric composite is composed of piezoelectric phase and matrix phase. In 1-3 piezoelectric composite, piezoelectric rods are perpendicularly embedded in a viscoelastic matrix and the two phases are mechanically connected in parallel. Since the piezoelectric phase is connected in one direction and the matrix phase is connected in three directions, the composite is called 1-3 type according to the naming rule. The one-direction connectivity of piezoelectric rods is a key factor deciding high piezoelectricity of the composite. The presence of a viscoelastic matrix with low density and low stiffness contributes to a good flexibility and a low acoustic impedance of the composite.

In recent years, a novel 1-3-2 piezoelectric composite is proposed to overcome the instability of 1-3 composite structure and at the same time inherits the advantages of 1-3 composite \cite{Wan08,Li09,Qin10,Li13,Sak10,Sak12,Zho12,Xu14,Lu14}. 1-3-2 piezoelectric composite consists of two layers which are mechanically connected in series. One layer is 1-3 composite and the other layer is a piezoelectric ceramic base layer. The ``2'' in ``1-3-2'' means that the piezoelectric phase in the ceramic base layer is connected in two directions. The transverse ceramic base and the longitudinal piezoelectric rods form a framework that can support the whole composite structure. This series and parallel combination in structure greatly increases the stability of the composite. Therefore, the composite does not deform easily in the fabrication process and is free from the influence of external mechanical impact and the environment temperature change. Since 1-3 composite is a component of 1-3-2 composite, 1-3-2 composite has characteristics similar to 1-3 composite and can also be effectively used in ultrasonic transducers.

Wang et al. \cite{Wan08,Li09,Qin10,Li13} have estimated the effective properties of 1-3-2 composite based on linear piezoelectric theory and uniform field theory. They determined the influence of piezoelectric phase volume fraction and composite aspect (thickness/width) on the resonance characteristic by finite element method. They also manufactured 1-3-2 composite samples to measure the parameters, such as hydrostatic pressure stability, temperature stability, voltage response and receiving voltage sensitivity. Sakthivel and Arockiarajan \cite{Sak10,Sak12} have developed a micromechanics-based analytical model to capture the effective thermo-electro-mechanical response of 1-3-2 composite and predicted the electro-elastic constants. Zhou et~al. \cite{Zho12} have fabricated 1-3-2 type multi-element piezoelectric composite samples and measured their properties. The results showed that 1-3-2 composite has a high thickness electromechanical coupling factor and a wide bandwidth of pulse-echo signal. Xu et~al. \cite{Xu14} have studied the influence of cement matrix on properties of 1-3-2 piezoelectric composite. The involved properties are as follows: relative dielectric factor, dielectric loss, planar electromechanical coupling coefficient, mechanical quality factor, acoustic impedance and thickness electromechanical coupling coefficient. Luan et~al. \cite{Lu14} have calculated the relationship between the thickness of 1-3-2 type piezoelectric composite and its resonant frequency of admittance characteristics by finite element analysis software ANSYS.

More recently, 2-1-3 piezoelectric composite was proposed to loosen the manufacturing fineness requirement of 1-3-2 composite under a so-called ``clamped underside boundary condition'' in practical application \cite{Sun16}. The ``clamped underside boundary condition'' simulates the situation when a piezoelectric transducer is designed to be a coating and is covered on a main structure with large volume and mass. Sun et~al. \cite{Sun16} analyzed the dependence of electric-elastic properties of 1-3-2 and 2-1-3 composites on the aspect ratio by finite element method and concluded that the fineness of 30\%-volume faction 2-1-3 composite can be reduced to 2.54\% of that of 1-3-2 composite. 2-1-3 composite is a revised version of 1-3-2 and can be obtained by turning over a 1-3-2 composite. Namely, 2-1-3 composite also consists of 1-3 composite and a piezoelectric ceramic layer. Although the piezoelectric ceramic layer is not treated as a base under 1-3 composite, the covering on 1-3 composite and, therefore, the composite is referred to as 2-1-3 type. 2-1-3 composites possess a good temperature and mechanical stability, the same as 1-3-2 composite. Since the performance of 1-3-2 composite is sensitive to an increase of the lateral spatial scale of piezoelectric rods, it is necessary to ensure a high manufacturing fineness requirement. However, this requirement of 2-1-3 composite can be relaxed a lot while retaining the needed performance.

In conclusion, 1-3, 1-3-2 and 2-1-3 piezoelectric composites have a common component structure, so there are some similarities in their properties. They are proposed in different stages for solving different problems. 1-3 composite is studied to obtain a more suitable piezoelectric transducer material. Then, 1-3-2 composite is proposed in order to enhance the mechanical stability. Lately 2-1-3 composite is designed to reduce the manufacturing difficulty under a specific application condition. Hence, the three have respective advantages and limitations and it is quite necessary to systematically compare their material properties and make a comprehensive summary.

The lateral spatial scale of piezoelectric rods in 1-3, 1-3-2 and 2-1-3 composites largely influence the electro-elastic properties of the composites \cite{Hos91}. From this perspective, the composites can be divided into two types, namely ``fine-scaled composite'' and ``coarse-scaled composite''. It is necessary to make a distinction between these two types and study them in different ways. Since a fine-scaled composite can be treated as an effective homogeneous medium, it is reasonable to apply some assumptions and conduct theoretical derivation \cite{Li09,Qin10,Li13,Sak10,Sak12,Smi91,Smi93,Cha89}. However, the movements of the two phases in coarse-scaled composite are not synchronous and, hence, it is a fairly complex problem to carry out theoretical modelling and derivation. Finite element method (FEM) is a more suitable choice to do this research. During FEM modelling, a key step that needs special attention is loading boundary conditions. An improper handing has been found in Madhusudhana Rao's work \cite{Rao10}. In his paper, the perpendicular displacements on all nodes on the surface of the composite unit cell are restrained in order to be the same. This leads to equal vertical strains of both piezoelectric phase and matrix phase, which is definitely not in accord with the fact of coarse-scaled composite. Therefore, the obtained results of the coupling coefficient are unreliable. A detailed discussion on this problem is in \cite{Sun14}.

This paper involves continuing and further research on 2-1-3 piezoelectric composite serving as a kind of receiving transducer material. The ``clamped underside boundary condition'' in the previous research work of 2-1-3 composite is actually set according to the application condition of a receiving transducer. A comparative study on fine-scaled 1-3, 1-3-2 and 2-1-3 composite transducers is conducted first. Then, a comparative study on coarse-scaled 1-3, 1-3-2 and 2-1-3 composite transducers is presented. The concerned performance parameters are longitudinal velocity, thickness electromechanical coupling coefficient and voltage receiving sensitivity. It is concluded that 2-1-3 composite is of higher piezoelectricity and a better voltage receiving sensitivity performance. At last, the essential causes are elaborated.

\section{Modelling}
\subsection{Model of 1-3 composite transducer}
\begin{figure}[!b]
\centerline{\includegraphics[width=0.5\textwidth]{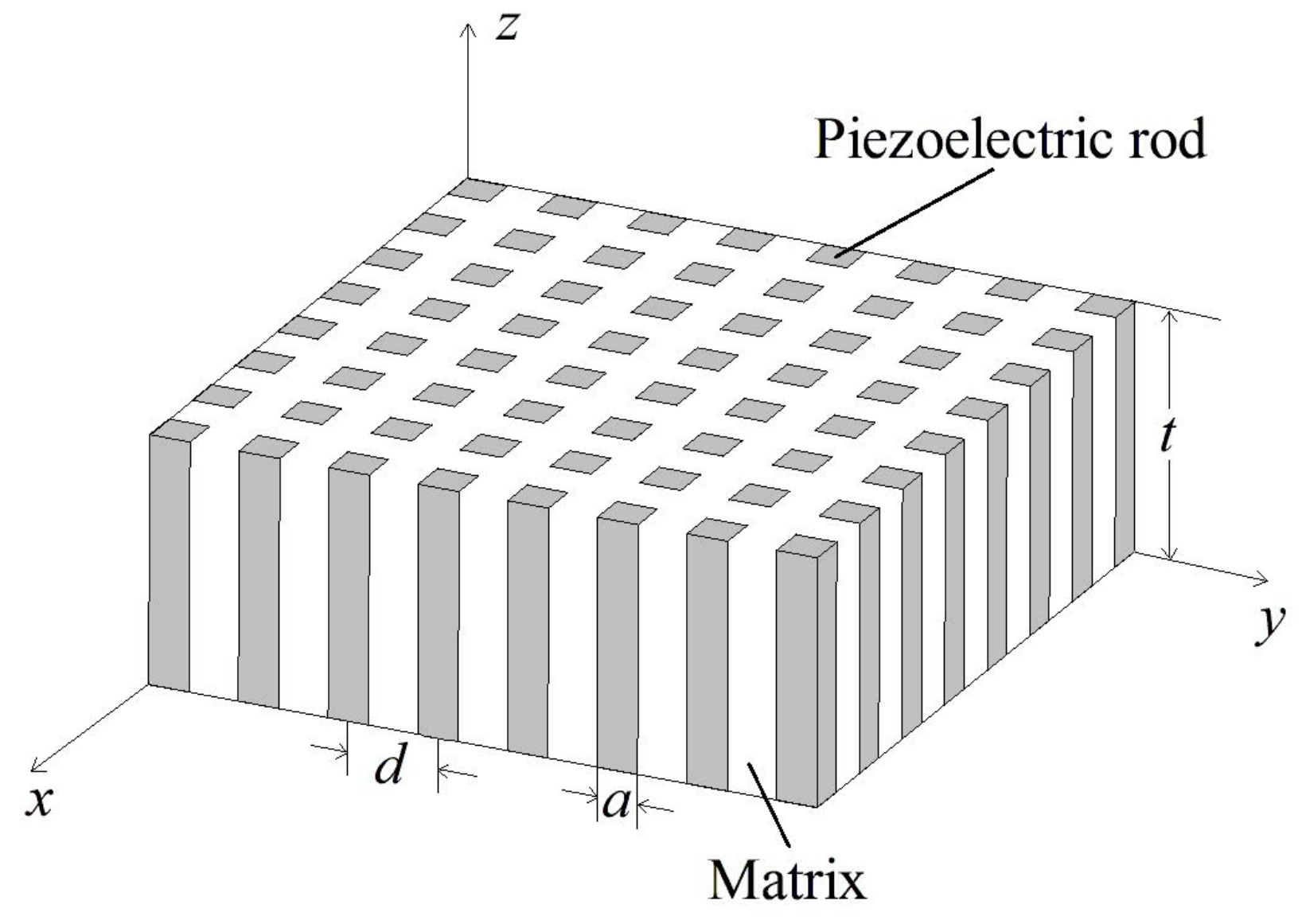}}
\caption{(Colour online) 1-3 piezoelectric composite transducer.} 
\label{figure-1}
\end{figure}

In this paper, 1-3 piezoelectric composite transducer is modelled with cubic piezoelectric rods that are perpendicularly embedded across the thickness ($z$-direction) of a viscoelastic matrix in the coordinate system $(x,y,z)$ as shown in figure \ref{figure-1}. The piezoelectric rods are poled in $z$-direction and the matrix is assumed to be piezoelectrically inactive and homogeneously isotropic. The underside of the composite is on $x{-}y$~plane and is clamped in its normal direction. The top surface is the stress surface.

Let $t$ represent the thickness of the composite and $d$ represent the lateral periodicity of piezoelectric rods. The ratio of $d$ to $t$ can directly influence the properties of 1-3 composite transducer. It is called the aspect ratio and is defined as:

\begin{equation}
\alpha_{(1\text{-}3)}=\frac{d}{t} \label{equation-1}.
\end{equation}

Let $a$ represent the side length of the cross-section of the cubic piezoelectric rod. Then, the volume fraction of the piezoelectric phase (VFP) can be expressed as follows:

\begin{equation}
VPF_{(1\text{-}3)}=\frac{a^2}{d^2} \label{equation-2}.
\end{equation}

\subsection{Model of 1-3-2 composite transducer}
\begin{figure}[htb]
\centerline{\includegraphics[width=0.55\textwidth]{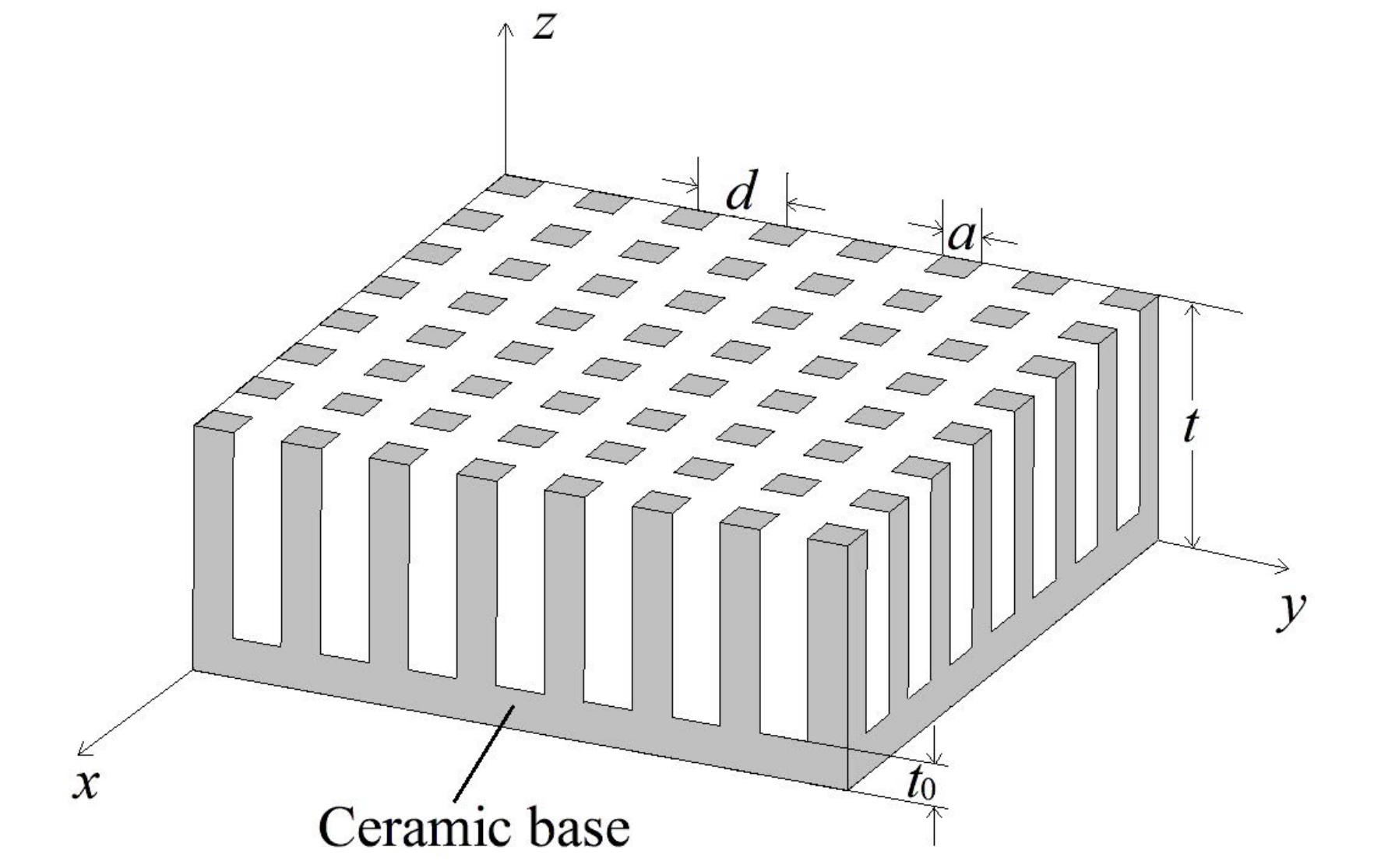}}
\caption{(Colour online) 1-3-2 piezoelectric composite transducer.} 
\label{figure-2}
\end{figure}

1-3-2 piezoelectric composite is a modified version of 1-3 piezoelectric composite as shown in figure~\ref{figure-2}, which consists of 1-3 composite and a ceramic base layer. The two parts are connected in series. The underside of the ceramic base is set as the underside of the composite and is clamped in its normal direction. The top surface of 1-3 composite is the stress surface.

Let $t$ represent the thickness of 1-3-2 composite and $d$ represent the lateral periodicity of piezoelectric rods. The aspect ratio is

\begin{equation}
\alpha_{(1\text{-}3\text{-}2)}=\frac{d}{t} \label{equation-3}.
\end{equation}

Let $t_0$ represent the thickness of the ceramic base layer. Therefore, the thickness of 1-3 composite part is $t-t_0$. The ratio of $t_0$ to $t$ is defined as  follows:

\begin{equation}
\beta_{(1\text{-}3\text{-}2)}=\frac{t_0}{t} \label{equation-4}.
\end{equation}

It can be derived that the volume fraction of the piezoelectric phase (VFP) in 1-3-2 composite is

\begin{equation}
VPF_{(1\text{-}3\text{-}2)}=\beta_{(1\text{-}3\text{-}2)}+\frac{a^2}{d^2}\left(1-\beta_{(1\text{-}3\text{-}2)}\right) \label{equation-5}.
\end{equation}

\subsection{Model of 2-1-3 composite transducer}
\begin{figure}[!t]
\centerline{\includegraphics[width=0.55\textwidth]{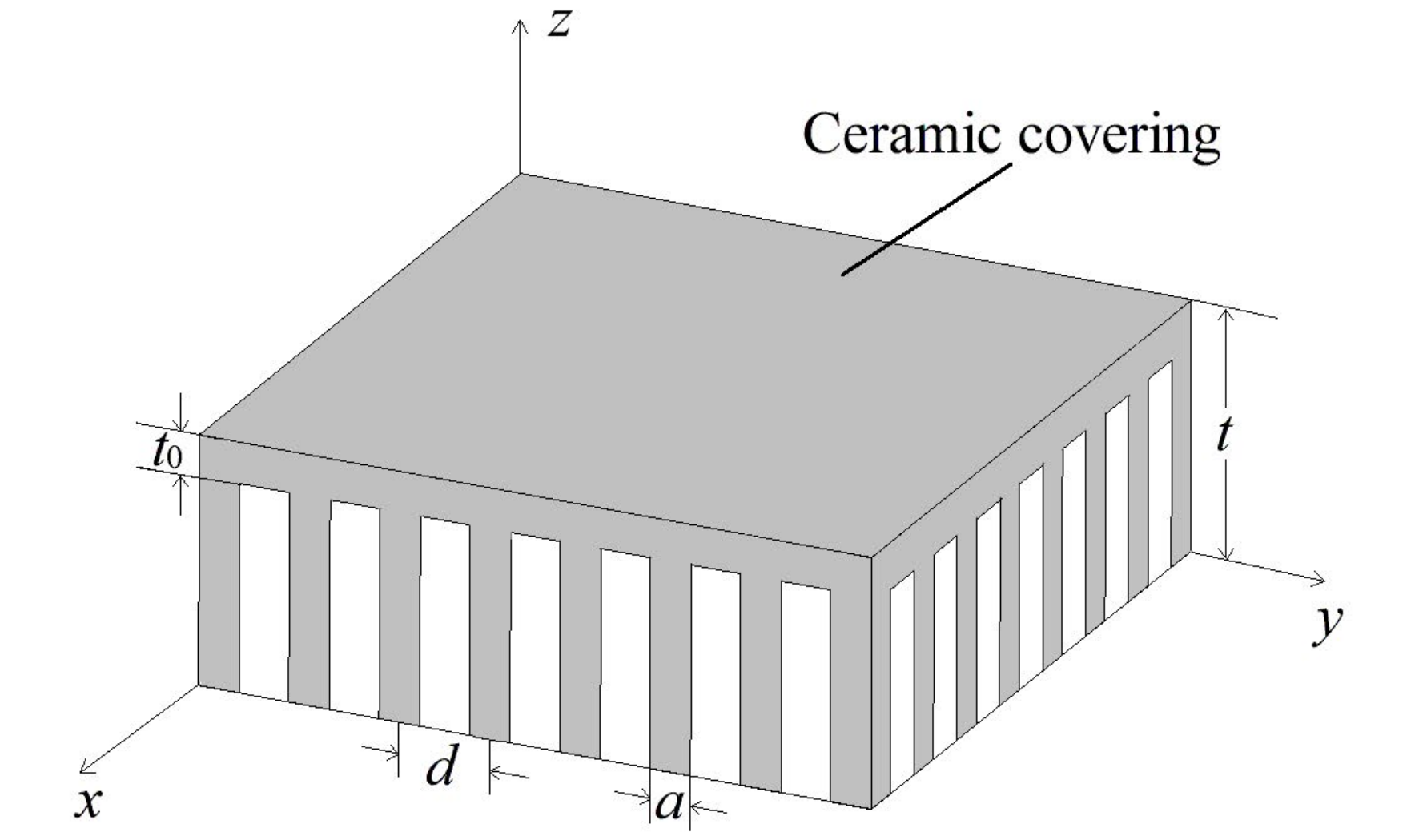}}
\caption{(Colour online) 2-1-3 piezoelectric composite transducer.} 
\label{figure-3}
\end{figure}

If the ceramic base layer of 1-3-2 composite is moved from the bottom to the top of 1-3 composite, a new revised version of 1-3-2 composite is generated as shown in figure~\ref{figure-3}, which is named ``2-1-3 composite'' according to the naming rules of piezoelectric composite. The surface of the ceramic covering layer is set as the top surface of 2-1-3 composite and is also the stress surface. The underside of 1-3 composite is set as the underside of the composite and is clamped in its normal direction.

For 2-1-3 composite, $t_0$ represents the thickness of the ceramic covering layer. The definitions of the dimension parameters including $a$, $d$, $t$, $\alpha$ and $\beta$ of 2-1-3 composite are the same as those of 1-3-2 composite. 

The aspect ratio is

\begin{equation}
\alpha_{(2\text{-}1\text{-}3)}=\frac{d}{t} \label{equation-6}.
\end{equation}

The ratio $\beta$ is

\begin{equation}
\beta_{(2\text{-}1\text{-}3)}=\frac{t_0}{t} \label{equation-7}.
\end{equation}

The volume fraction of the piezoelectric phase (VFP) in 2-1-3 composite is

\begin{equation}
VPF_{(2\text{-}1\text{-}3)}=\beta_{(2\text{-}1\text{-}3)}+\frac{a^2}{d^2}\left(1-\beta_{(2\text{-}1\text{-}3)}\right) \label{equation-8}.
\end{equation}

It can be concluded that if  $\beta=0$, which means the ceramic base does not exist for 1-3-2 composite or the ceramic covering does not exist for 2-1-3 composite, equations~(\ref{equation-5}) and (\ref{equation-8}) will turn into equation~(\ref{equation-2}), namely the volume fraction of the piezoelectric phase in 1-3 composite.

\section{Fine-scaled composite transducer}
When the piezoelectric rods in the composites are of fine lateral spatial scale, namely the aspect ratio $\alpha$ is very small, 1-3 composite itself and 1-3 composite part in 1-3-2 and 2-1-3 composites can all be treated as effective homogeneous media. This kind of a composite is called ``fine-scaled composite''. The parallel and series theory has been applied to theoretically model the fine-scaled 1-3 \cite{Smi91}, 1-3-2 \cite{Sak10} and 2-1-3 \cite{Sun16} piezoelectric composites. These models will be briefly reviewed as follows.

\subsection{Fine-scaled 1-3 composite}
If strain $S$ and electric field $E$ are chosen as independent coordinates for the analysis, the constitutive relations of the matrix phase and the piezoelectric phase can be simplified as equations~(\ref{equation-9}) and (\ref{equation-10}).

\begin{equation}       
\left(                 
  \begin{array}{c} 
 \vspace{1mm}  
    T_x^\textrm{m}\\
 \vspace{1mm}  
    T_z^\textrm{m}\\  
    D_z^\textrm{m}    
  \end{array}
\right)=
\left(
  \begin{array}{ccc}
 \vspace{1mm}
    c_{11}^\textrm{m}+c_{12}^\textrm{m} & c_{12}^\textrm{m} & 0\\
 \vspace{1mm}
    2c_{12}^\textrm{m} & c_{11}^\textrm{m} & 0\\
    0 & 0 & \epsilon_x^\textrm{m}
  \end{array}
\right)=
\left(
 \begin{array}{c}
 \vspace{1mm}
    S_x^\textrm{m}\\
 \vspace{1mm}
    S_z^\textrm{m}\\
    E_z^\textrm{m}
  \end{array}
\right), \label{equation-9}
\end{equation}

\begin{equation}       
\left(                 
  \begin{array}{c} 
 \vspace{1mm}  
    T_x^\textrm{p}\\
 \vspace{1mm}  
    T_z^\textrm{p}\\  
    D_z^\textrm{p}    
  \end{array}
\right)=
\left(
  \begin{array}{ccc}
 \vspace{1mm}
    c_{11}^E+c_{12}^E & c_{13}^E & -e_{31}\\
 \vspace{1mm}
    2c_{13}^E & c_{33}^E & -e_{33}\\
    2e_{31} & e_{33} & \epsilon_{33}^S
  \end{array}
\right)=
\left(
 \begin{array}{c}
 \vspace{1mm}
    S_x^\textrm{p}\\
 \vspace{1mm}
    S_z^\textrm{p}\\
    E_z^\textrm{p}
  \end{array}
\right). \label{equation-10}
\end{equation}

Here, superscript ``m'' stands for the matrix and ``p'' means the piezoelectric rods. In 1-3 composite, the piezoelectric phase and the matrix phase are connected in parallel. Several approximations indicating the relationship between the two phases are listed as follows. The involved variables are strain $S$, electric field $E$, stress $T$ and electrical displacement $D$.
\begin{eqnarray}
S_z^\textrm{p}=S_z^\textrm{m}=S_z^{1\text{-}3}, \label{equation-11} \\
E_z^\textrm{p}=E_z^\textrm{m}=E_z^{1\text{-}3}, \label{equation-12} \\
T_x^\textrm{p}=T_x^\textrm{m}=T_x^{1\text{-}3}, \label{equation-13} 
\end{eqnarray}
\begin{eqnarray}
S_x^{1\text{-}3}=\mu S_x^\textrm{p}+(1-\mu)S_x^\textrm{m}=0\,, \label{equation-14} \\
T_z^{1\text{-}3}=\mu T_z^\textrm{p}+(1-\mu)T_z^\textrm{m}, \label{equation-15} \\
D_z^{1\text{-}3}=\mu D_z^\textrm{p}+(1-\mu)D_z^\textrm{m}. \label{equation-16}
\end{eqnarray}

Here, $\mu$ is the volume fraction of the piezoelectric phase and $1-\mu$ is that of the matrix phase. The elastic constant $c_{33(1\text{-}3)}^E$ and $c_{33(1\text{-}3)}^D$, the dielectric constant $\epsilon_{33(1\text{-}3)}^S$ and the piezoelectric constant $e_{33(1\text{-}3)}$ of 1-3 composite can be obtained as equations (\ref{equation-17})--(\ref{equation-20}) \cite{Smi91}.
\begin{equation}
c_{33(1\text{-}3)}^E=\mu \Bigg[c_{33}^E-\frac{2(1-\mu)\big(c_{13}^E-c_{12}^\textrm{m}\big)^2}{\mu \big(c_{12}^\textrm{m}+c_{11}^\textrm{m}\big)+(1-\mu)\big(c_{12}^E+c_{11}^E\big)}\Bigg]+(1-\mu)c_{11}^\textrm{m}\,, \label{equation-17}
\end{equation}
\begin{equation}
\epsilon_{33(1\text{-}3)}^S=\mu \Bigg[\epsilon_{33}^S+\frac{2(1-\mu)e_{31}^2}{\mu \big(c_{12}^\textrm{m}+c_{11}^\textrm{m}\big)+(1-\mu)\big(c_{12}^E+c_{11}^E)}\Bigg]+(1-\mu)\epsilon_x^\textrm{m}, \label{equation-18}
\end{equation}
\begin{equation}
e_{33(1\text{-}3)}=\mu \Bigg[e_{33}-\frac{2(1-\mu)e_{31}\big(c_{13}^E-c_{12}^\textrm{m}\big)}{\mu \big(c_{12}^\textrm{m}+c_{11}^\textrm{m}\big)+(1-\mu)\big(c_{12}^E+c_{11}^E\big)}\Bigg], \label{equation-19}
\end{equation}
\begin{equation}
c_{33(1\text{-}3)}^D=c_{33(1\text{-}3)}^E+\frac{e_{33(1\text{-}3)}^2}{\epsilon_{33(1\text{-}3)}^S}\,. \label{equation-20}
\end{equation}

\subsection{Fine-scaled 1-3-2 and 2-1-3 composites}
According to \cite{Sun16}, both fine-scaled 1-3-2 composite and fine-scaled 2-1-3 composite can be considered as a conventional laminated composite with two layers. One layer is a fine-scaled 1-3 composite and the other layer is a ceramic layer. The constitutive relations of the two layers can be simplified as equations~(\ref{equation-21}) and (\ref{equation-22}), respectively.

\begin{equation}       
\left(                 
  \begin{array}{c} 
  \vspace{1mm}  
    T_x^\textrm{cl}\\
  \vspace{1mm}  
    T_z^\textrm{cl}\\  
    D_z^\textrm{cl}    
  \end{array}
\right)=
\left(
 \begin{array}{ccc}
 \vspace{1mm}
    c_{11}^E+c_{12}^E & c_{13}^E & -e_{31}\\ 
 \vspace{1mm}
     2c_{13}^E & c_{33}^E & -e_{33}\\ 
    2e_{31} & e_{33} & \epsilon_{33}^S
  \end{array}
\right)=
\left(
 \begin{array}{c}
 \vspace{1mm}
    S_x^\textrm{cl}\\
 \vspace{1mm}
    S_z^\textrm{cl}\\
    E_z^\textrm{cl}
  \end{array}
\right), \label{equation-21}
\end{equation}

\begin{equation}       
\left(                 
  \begin{array}{c}
 \vspace{1mm}   
    T_x^{1\text{-}3}\\
 \vspace{1mm}  
    T_z^{1\text{-}3}\\  
    D_z^{1\text{-}3}    
  \end{array}
\right)=
\left(
  \begin{array}{ccc}
 \vspace{1mm}
    c_{11(1\text{-}3)}^E+c_{12(1\text{-}3)}^E & c_{13(1\text{-}3)}^E & -e_{31(1\text{-}3)}\\
 \vspace{1mm}
    2c_{13(1\text{-}3)}^E & c_{33(1\text{-}3)}^E & -e_{33(1\text{-}3)}\\
    2e_{31(1\text{-}3)} & e_{33(1\text{-}3)} & \epsilon_{33(1\text{-}3)}^S
  \end{array}
\right)=
\left(
 \begin{array}{c}
 \vspace{1mm}
    S_x^{1\text{-}3}\\
 \vspace{1mm}
    S_z^{1\text{-}3}\\
    E_z^{1\text{-}3}
  \end{array}
\right). \label{equation-22}
\end{equation}

Here, ``cl'' means the ceramic layer and ``1-3'' represents the 1-3 composite layer. In 1-3-2 or 2-1-3 composite, the two layers are connected in series. Several approximations indicating the relationship between the two layers are listed as follows:
\begin{eqnarray}
&T_z^\textrm{cl}=T_z^{1\text{-}3}=T_z^{1\text{-}3\text{-}2},& \label{equation-23} \\
&D_z^\textrm{cl}=D_z^{1\text{-}3}=D_z^{1\text{-}3\text{-}2},& \label{equation-24} \\
&S_x^\textrm{cl}=S_x^{1\text{-}3}=S_x^{1\text{-}3\text{-}2},& \label{equation-25} \\
&T_x^{1\text{-}3\text{-}2}=\beta T_x^\textrm{cl}+(1-\beta)T_x^{1\text{-}3},& \label{equation-26} \\
&S_z^{1\text{-}3\text{-}2}=\beta S_z^\textrm{cl}+(1-\beta)S_z^{1\text{-}3},& \\ 
&E_z^{1\text{-}3\text{-}2}=\beta E_z^\textrm{cl}+(1-\beta)E_z^{1\text{-}3}.&  
\end{eqnarray}

Here, $\beta$ and $1-\beta$ are volume fractions of the ceramic layer and 1-3 composite layer in 1-3-2 or 2-1-3 composite. The electro-elastic constants of fine-scaled 1-3-2 or 2-1-3 piezoelectric composite can be derived as equations~(\ref{equation-29})--(\ref{equation-35}) \cite{Sak10,Sun16}.

\begin{equation}
c_{33(1\text{-}3\text{-}2)}^E=c_{33(1\text{-}3)}^E\big(\overline{K}_1c_{33}^E+\overline{K}_2e_{33}\big)-e_{33(1\text{-}3)}\big(-\overline{K}_2c_{33}^E+\overline{K}_3e_{33}\big), \label{equation-29}
\end{equation}
\begin{equation}
e_{33(1\text{-}3\text{-}2)}=e_{33(1\text{-}3)}\big(\overline{K}_1c_{33}^E+\overline{K}_2e_{33}\big)+\epsilon_{33(1\text{-}3)}^S
\big(-\overline{K}_2c_{33}^E+\overline{K}_3e_{33}\big), \label{equation-30}
\end{equation}
\begin{equation}
\epsilon_{33(1\text{-}3\text{-}2)}^S=e_{33(1\text{-}3)}\big(-\overline{K}_1e_{33}+\overline{K}_2\epsilon_{33}^S\big)+\epsilon_{33(1\text{-}3)}^S
\big(\overline{K}_2e_{33}+\overline{K}_3\epsilon_{33}^S\big), \label{equation-31}
\end{equation}
\begin{equation}
\overline{K}_1=\frac{\epsilon_{33(1\text{-}3)}^S\beta+\epsilon_{33}^S(1-\beta)}{\big(\epsilon_{33(1\text{-}3)}^S\beta+\epsilon_{33}^S(1-\beta)\big)\big(c_{33(1\text{-}3)}^E\beta+c_{33}^E(1-\beta)\big)+\big(e_{33(1\text{-}3)}\beta+e_{33}(1-\beta)\big)^2}\,, \label{equation-32}
\end{equation}
\begin{equation}
\overline{K}_2=\frac{e_{33(1\text{-}3)}\beta+e_{33}(1-\beta)}{\big(\epsilon_{33(1\text{-}3)}^S\beta+\epsilon_{33}^S(1-\beta)\big)\big(c_{33(1\text{-}3)}^E\beta+c_{33}^E(1-\beta)\big)+\big(e_{33(1\text{-}3)}\beta+e_{33}(1-\beta)\big)^2}\,, \label{equation-33}
\end{equation}
\begin{equation}
\overline{K}_3=\frac{c_{33(1\text{-}3)}^E\beta+c_{33}^E(1-\beta)}{\big(\epsilon_{33(1\text{-}3)}^S\beta+\epsilon_{33}^S(1-\beta)\big)\big(c_{33(1\text{-}3)}^E\beta+c_{33}^E(1-\beta)\big)+\big(e_{33(1\text{-}3)}\beta+e_{33}(1-\beta)\big)^2}\,, \label{equation-34}
\end{equation}
\begin{equation}
c_{33(1\text{-}3\text{-}2)}^D=c_{33(1\text{-}3\text{-}2)}^E+\frac{e_{33(1\text{-}3\text{-}2)}^2}{\epsilon_{33(1\text{-}3\text{-}2)}^S}\,. \label{equation-35}
\end{equation}

\subsection{Performance comparison of fine-scaled composite transducers} \label{section3_3}
\begin{table}[!b]  
	\scriptsize
	\caption{Parameters of phase materials.}  
	\begin{center} 
		\renewcommand{\arraystretch}{1.5} 
		\begin{tabular}{c|c|c|c|c|c|c|c|c|c|c|c|c}  
			\hline \hline  
			\multirow{3}{*}{} & \multirow{1}{*}{Density} & \multicolumn{6}{|c|}{Elastic constants} & \multicolumn{3}{c}{Piezoelectric} & \multicolumn{2}{|c}{Relative dielectric}\\ 
			\multirow{3}{*}{} & \multirow{1}{*}{} & \multicolumn{6}{|c|}{} & \multicolumn{3}{c}{constants} & \multicolumn{2}{|c}{constant}\\ 
			\cline{2-13}  
			& $\rho$ & $c_{11}$ & $c_{12}$ & $c_{13}$ & $c_{33}$ & $c_{44}$ & $c_{66}$ & $e_{15}$ & $e_{31}$ & $e_{33}$ & $\epsilon_{11}/\epsilon_0$ & $\epsilon_{33}/\epsilon_0$  \\
			\hline
			Phase materials & \multirow{1}{*}{$\textrm{kg/m}^3$} & \multicolumn{6}{|c|}{$10^{10}\textrm{Pa}$} & \multicolumn{3}{|c|}{$\textrm{C/m}^2$} & \multicolumn{2}{|c}{$-$}\\  
			\cline{1-13}
			PZT-5H & 7500 &	12.6 & 7.95 & 8.41 & 11.7 & 2.3 & 2.35 & 17.0 & $-$6.5 & 23.3 & 1700 & 1470\\
			\cline{1-13}
			Rubber & 1003 & 0.23 &	0.22 & 0.22 & 0.23 & 0.005 &  0.005 & $-$ & $-$ & $-$ & 3 & 3\\
			\hline \hline  
		\end{tabular}  
	\end{center} 
	\label{table-1} 
\end{table}

It is clearly seen that the electro-elastic constants of fine-scaled 1-3, 1-3-2 and 2-1-3 composites are decided by the parameters and volume fractions of two phases and have nothing to do with the aspect ratio~$\alpha$. The longitudinal velocity $v_\textrm{l}$ and the thickness electromechanical coupling coefficient $k_\textrm{t}$ of fine-scaled 1-3, 1-3-2 and 2-1-3 composite transducers can be obtained from the above electro-elastic constants and are expressed as follows:
\begin{equation}
v_\textrm{l}=\sqrt{\frac{c_{33}^D}{\rho}}\,, \label{equation-36}
\end{equation}
\begin{equation}
\rho=\rho^\textrm{p}VFP+\rho^\textrm{m}(1-VFP)\,, \label{equation-37}
\end{equation}
\begin{equation}
k_{\textrm{t}}=\sqrt{1-\frac{c_{33}^E}{c_{33}^D}}\,. \label{equation-38}
\end{equation}

\begin{figure}[!t]
	\centerline{\includegraphics[width=0.51\textwidth]{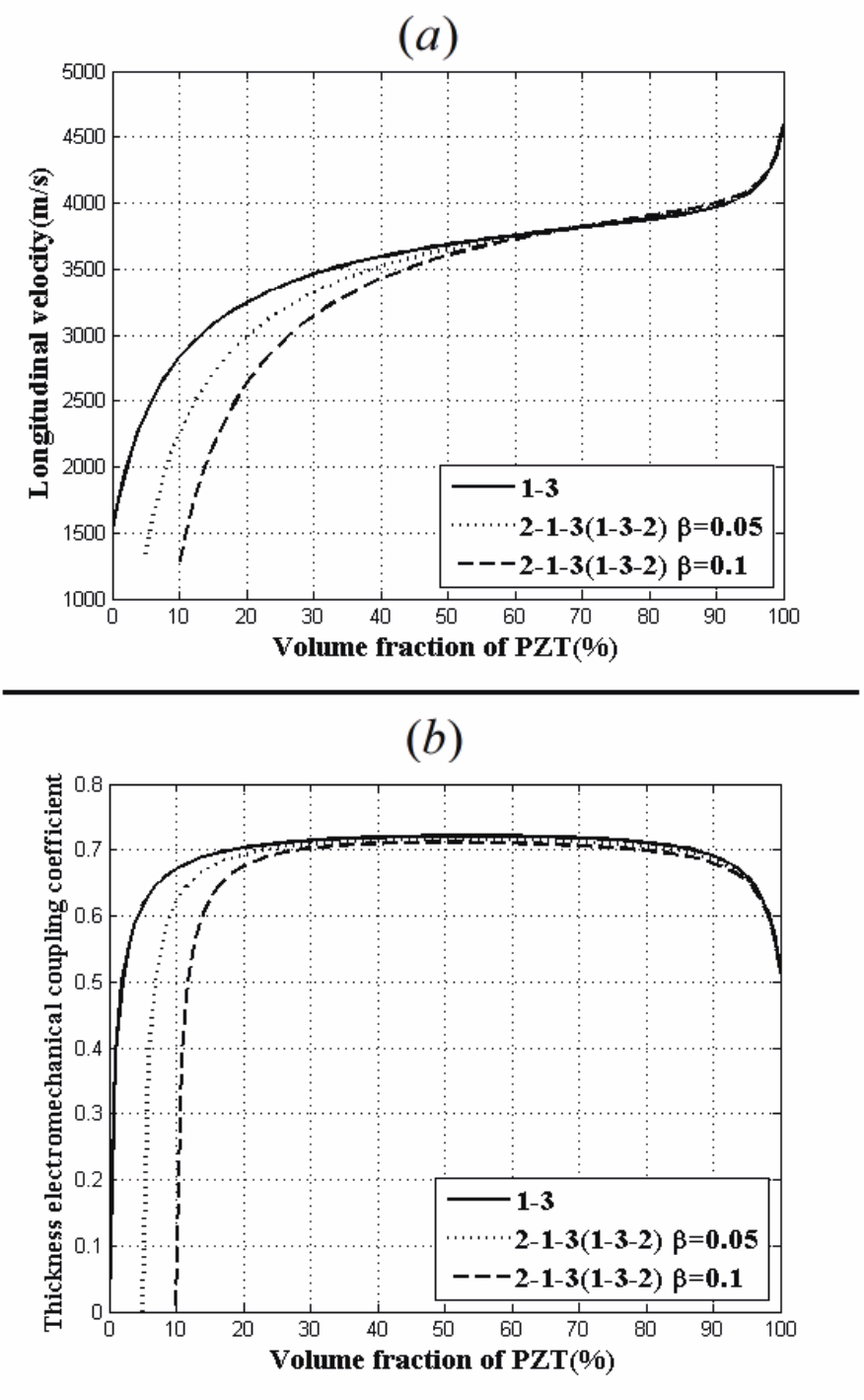}}
	\caption{The dependence of (a) longitudinal velocity, (b) thickness electromechanical coupling coefficient of 
		fine-scaled 1-3, 1-3-2 and 2-1-3 composites on volume fraction of PZT.} 
	\label{figure-4}
\end{figure}

In this paper, PZT-5H and natural rubber are adopted as the piezoelectric phase and the matrix phase. The parameters of the two materials are listed in table~\ref{table-1}. The dependences of the longitudinal velocity $v_{\textrm{l}}$ and the thickness electromechanical coupling coefficient $k_{\textrm{t}}$ of fine-scaled 1-3, 1-3-2 and 2-1-3 composites on the volume fraction of piezoelectric phase are shown in figure~\ref{figure-4}~(a) and (b). From the results shown in figure~\ref{figure-4} we can conclude that there is no difference between fine-scaled 1-3-2 composite and fine-scaled 2-1-3 composite in terms of electro-elastic properties. The variation of $v_{\textrm{l}}$ or $k_{\textrm{t}}$ of fine-scaled 1-3-2 or 2-1-3 composite is similar to that of fine-scaled 1-3 composite. Since high thickness electromechanical coupling coefficient, low longitudinal velocity and low density are necessary for practical applications of piezoelectric transducer, 30\% can be one of the proper selections of volume fraction of piezoelectric phase.

\section{Coarse-scaled composite transducer}
One of the essential approximations [equation (\ref{equation-11})] in the theoretical models of fine-scaled 1-3 composite and the 1-3 composite part in fine-scaled 1-3-2 or 2-1-3 composite is that the vertical strains are the same in both piezoelectric and matrix phases. This is hereinafter referred to as the ``same strain in both phases'' condition, which is a typical feature of fine-scaled composites.

\begin{figure}[!t]
	\centerline{\includegraphics[width=0.9\textwidth]{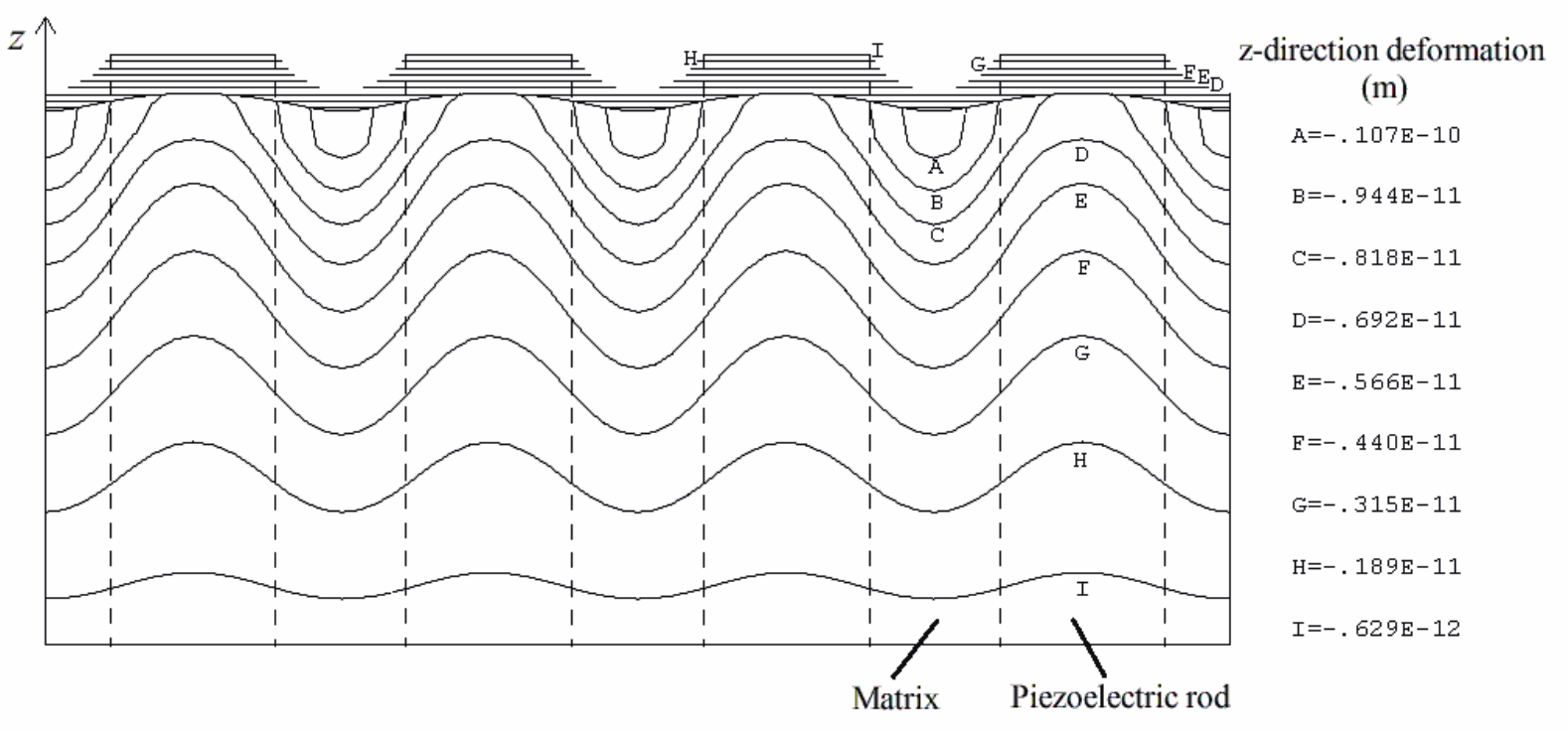}}
	\caption{Contour of $z$-direction deformation of 1-3 composite.} 
	\label{figure-5}
\end{figure} 

Obviously, it is better that the lateral spatial scale of piezoelectric rods can turn to much more coarse (namely the aspect ratio $\alpha$ turns to much larger) in order to reduce the complexity of the manufacture craft. What kind of changes the larger aspect ratio will bring to the composites are questions worth studying. However, when the lateral spatial scale becomes coarse to a certain extent, the above ``same strain in both phases'' condition cannot be met. Therefore, it is a task for finite element method (FEM) instead of the theoretical model to precisely quantify the performance parameters of coarse-scaled composites. Figure~\ref{figure-5} gives the contour of $z$-direction deformation displacement of 1-3 composite ($\alpha=0.5$) calculated by FEM when a pressure is applied on its surface. It is obvious that the matrix phase produces a larger compression deformation than the piezoelectric phase. 

\begin{figure}[!b]
\centerline{\includegraphics[width=0.78\textwidth]{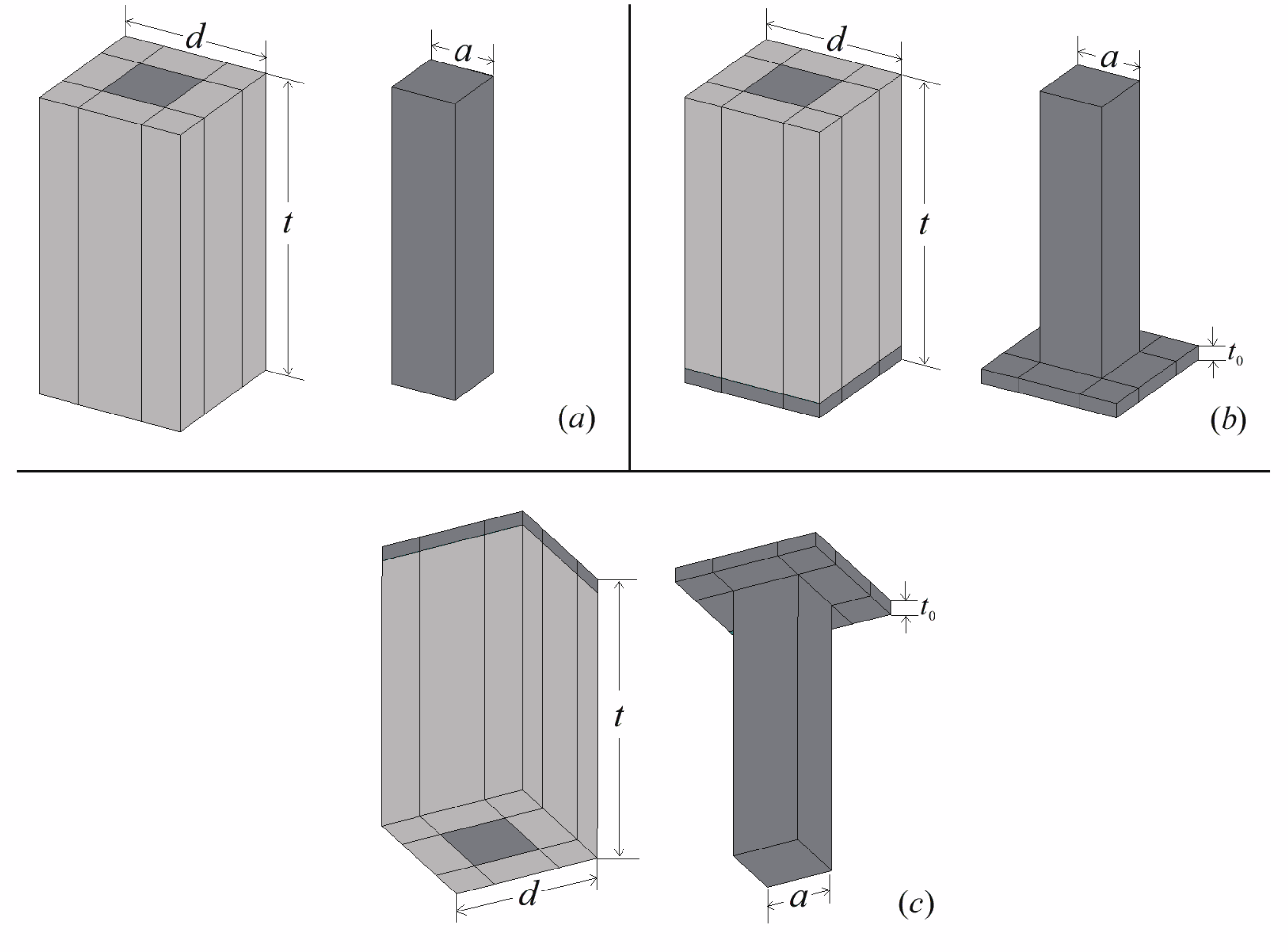}}
\vspace{-2mm}
\caption{(Colour online) Unit cell and piezoelectric phase of (a)1-3, (b)1-3-2 and (c) 2-1-3 piezoelectric composites.} 
\label{figure-6}
\end{figure}

\subsection{FEM Modelling}
Finite element simulation software ANSYS is used here to model 1-3, 1-3-2 and 2-1-3 composite transducers and to calculate electro-elastic constants $c_{33}^D$ and $c_{33}^E$ and voltage receiving sensitivity (VRS). Detailed finite element operations are briefly introduced, including preprocessor, modelling, meshing, boundary conditions and loads, solution and postprocessor.

Preprocessor: The element types of piezoelectric phase and matrix phase are both Solid5. In material props part, it is needed to input elastic constant matrix $[c^E]$, dielectric constant matrix $[\epsilon^S]$, piezoelectric constant matrix $[e]$ of piezoelectric phase, elastic constant matrix $[c^\textrm{m}]$ and dielectric constant matrix $[\epsilon^\textrm{m}]$ of passive phase. The parameters of piezoelectric phase (PZT-5H) and matrix phase (nature rubber) are listed in table~\ref{table-1}.

Modelling: The structures of 1-3, 1-3-2 and 2-1-3 composites are all of periodicity. Therefore, one of the unit cells can represent the property of the whole composite since it contains main characteristics of the microstructure. The unit cells of 1-3, 1-3-2 and 2-1-3 composites are shown in figure \ref{figure-6}~(a), (b) and (c), respectively. The thickness of the unit cell is the same as that of the composite and the side length of the cross-section of the composite is equal to the lateral periodicity of piezoelectrc rods.

Meshing: As the element types of both phases are tetrahedron element Solid5 and the degrees of freedom are UX, UY, UZ and VOLT [KEYOPT(1) is set as 3], mapped meshing is adopted to divide the composite unit cell into a lot of hexahedron grid units.

\subsection{Electro-elastic constants calculation by FEM}
Boundary conditions and loads: The longitudinal velocity $v_{\textrm{l}}$ and the thickness electromechanical coupling coefficient $k_{\textrm{t}}$ are decided by the elastic constants $c_{33}^D$ and $c_{33}^E$ as seen in equations~(\ref{equation-36}), (\ref{equation-37}) and (\ref{equation-38}). $c_{33}^D$ and $c_{33}^E$ of 1-3, 1-3-2 and 2-1-3 composites can be calculated in ANSYS by applying proper boundary conditions and loads to the unit cell model according to the definitions of the constants.
$c_{33}^D$ and $c_{33}^E$ are the variation of the stress tensor component $T_z$ caused by one-unit change of the strain tensor component $S_z$. The superscript ``D'' means the ``constant electrical displacement'' condition and ``E'' means the ``constant electrical field'' condition. Their definitions are

\begin{equation}
c_{33}^{D}=\left(\frac{\partial T_z}{\partial S_z}\right)_{D}=\left(\frac{\Delta T_z}{\Delta S_z}\right)\bigg|_{{D}=0}\,,\label{equation-39}
\end{equation}
\begin{equation}
c_{33}^{E}=\left(\frac{\partial T_z}{\partial S_z}\right)_{E}=\left(\frac{\Delta T_z}{\Delta S_z}\right)\bigg|_{E=0}\,. \label{equation-40}
\end{equation}

\begin{figure}[!b]
\centerline{\includegraphics[width=0.4\textwidth]{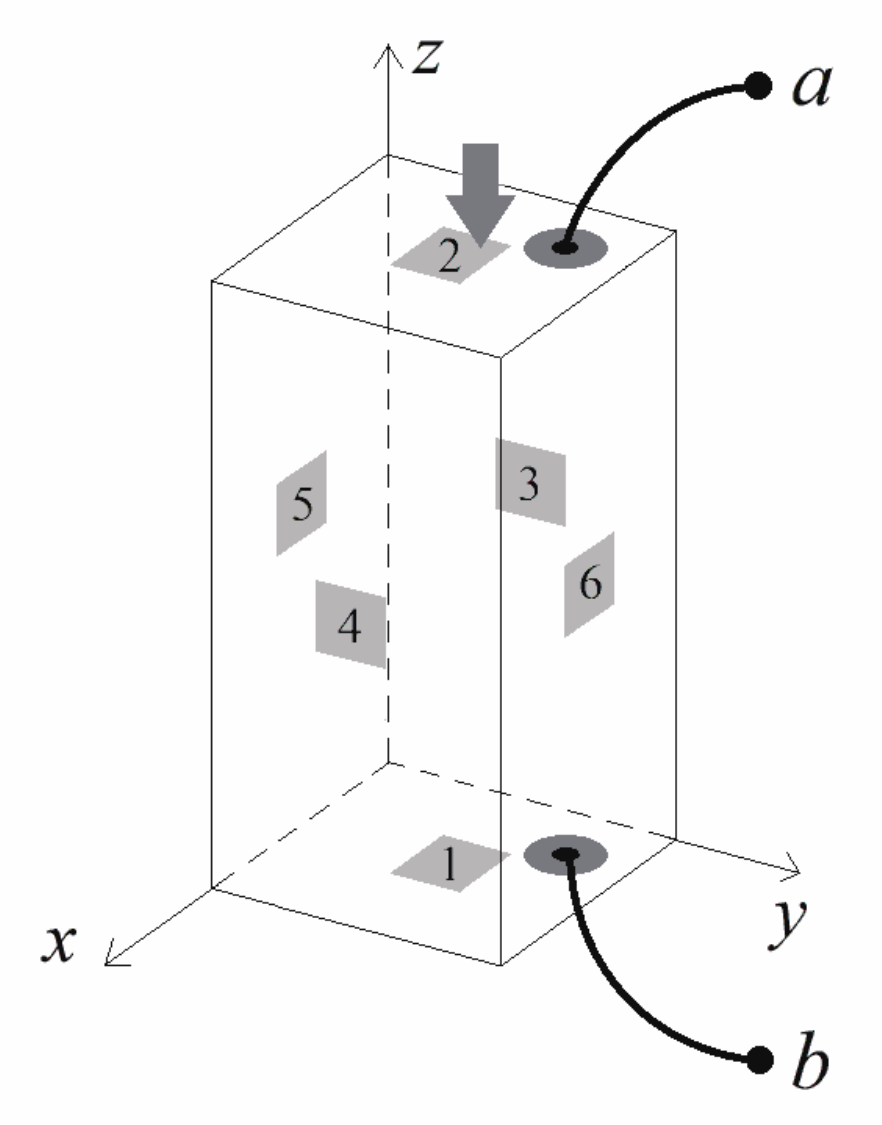}}
\caption{(Colour online) Boundary conditions and loads.} 
\label{figure-7}
\end{figure}

The profile of the composite unit cell in figure \ref{figure-7} is used to express boundary conditions and loads. To cause the change of the strain tensor component $S_z$, a pressure is loaded on plane 2 and plane 1 is fixed in $z$-direction. Planes 3-6 are also clamped in their normal directions in order to set strain tensors in other directions zero. The ``constant electrical field'' condition decides that planes 1 and 2 are short-circuited, namely electrodes  \textit{a} and \textit{b} are set zero potential. The ``constant electrical displacement'' condition decides that planes 1 and 2 are open-circuited.

Solution and postprocessor: Static analysis is carried out in solution part. In postprocessor part, the volume $(V_1,V_2,...,V_n)$, stress $(T_{z1},T_{z2},...,T_{zn})$ and strain $(S_{z1},S_{z2},...,S_{zn})$ of all the grid units are picked up to calculate the volumetrically weighted averages of the stress $\overline{T}_z$ and strain $\overline{S}_z$ and finally obtain $c_{33}$ using equation~(\ref{equation-41})
\begin{equation}
c_{33}=\frac{\overline{T}_z}{\overline{S}_z}=\frac{\sum_{j=1}^n {T_{zj}V_j}}{\sum_{j=1}^n {S_{zj}V_j}}\,. \label{equation-41}
\end{equation}

\subsection{Voltage receiving sensitivity calculation by FEM}
Boundary conditions and loads: Voltage receiving sensitivity (VRS) is a function of frequency and is one of the main parameters that can judge the working performance of a transducer. When a pressure $P_\textrm{a}$ is applied on the surface of a piezoelectric transducer, a voltage $V_\textrm{a}$ will emerge on the surface because of piezoelectric effect. Voltage receiving sensitivity is defined as the ratio of $V_\textrm{a}$ and $P_\textrm{a}$\,,

\begin{equation}
M=\frac{V_\textrm{a}}{P_\textrm{a}}. \label{equation-42}
\end{equation}

 If a reference value is set as $M_\textrm{r}=1\text V/$\textmu{}Pa, the voltage receiving sensitivity in dB is

\begin{equation}
M_{eL}=20\lg\frac{M}{M_\text{r}}=20\lg M-120~(\text{dB}). \label{equation-43}
\end{equation}

According to the definition, corresponding boundary conditions and loads are applied to the unit cell model in figure~\ref{figure-7}. Plane 1 is fixed in $z$-direction and planes 3--6 are clamped in their normal directions. A pressure is applied on plane~2 and the generated voltage will be picked up on the electrode \textit{a} on plane~2. The electrode \textit{b} is set zero potential.

Solution and postprocessor: In the solution part, harmonic response analysis is carried out since the wanted parameter is a function of frequency. The frequency range is set from 0 Hz to 200 kHz, which covers the first-order resonant frequency and first-order anti-resonant frequency of the composite transducer. In the postprocessor part, the voltage $V_\text{a}$ on the surface of the composite transducer is picked up and then the voltage receiving sensitivity is calculated.

\subsection{Comparison of longitudinal velocity}
\begin{figure}[!t]
\centerline{\includegraphics[width=0.9\textwidth]{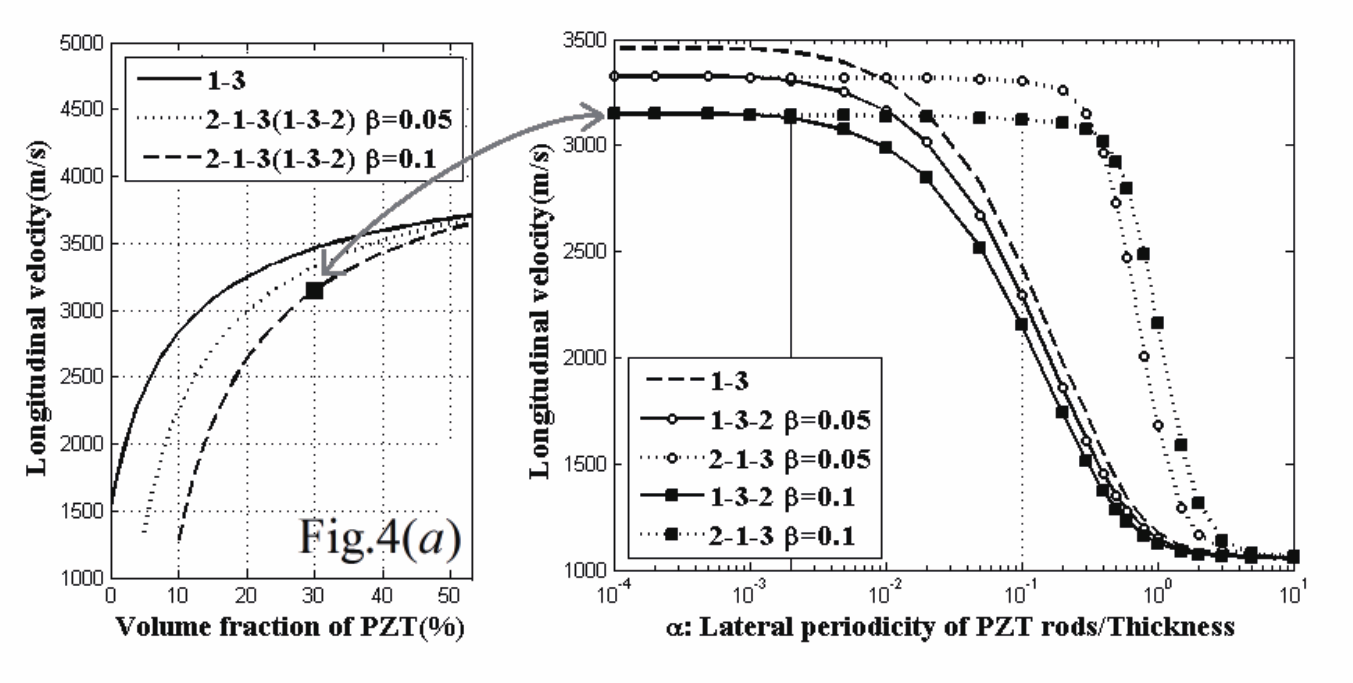}}
\caption{(Colour online) The dependence of longitudinal velocity of 1-3, 1-3-2 and 2-1-3 composites $(VFP=30\%)$ on the ratio $\alpha$.} 
\label{figure-8}
\end{figure}

In section~\ref{section3_3}, it is mentioned that 30\% can be one of the proper selections of volume fraction of piezoelectric phase. Therefore, VFPs of 1-3, 1-3-2 and 2-1-3 composites are all set as 30\% in the following calculation and analysis. The dependence of the longitudinal velocity $v_{\textrm{l}}$ on the ratio $\alpha$ is calculated and the result is shown in figure~\ref{figure-8}. 

The left-hand part of figure \ref{figure-8} is figure \ref{figure-4}~(a): the dependence of longitudinal velocity of fine-scaled 1-3, 1-3-2 and 2-1-3 composites on volume fraction of PZT. The solid square refers to the longitudinal velocity of 30\%-VFP fine-scaled 1-3-2 or 2-1-3 composite $(\beta=0.1)$. In the right-hand part of figure~\ref{figure-8}, when $\alpha$ is $10^{-4}$, which can be considered as small enough, the composites are of fine lateral spatial scale. The solid square here ($\alpha=10^{-4}$ and $\beta=0.1$) just corresponds to the solid square in figure~\ref{figure-4}~(a). This can prove the correctness of the calculation by FEM. 

For 1-3-2 and 2-1-3 composites $(\beta=0.1)$, the values of $\alpha$ at which the longitudinal velocity decreases by 1\% of their own maxima (when $\alpha=10^{-4}$) are called the critical values of $\alpha$, which are rounded up and are marked with a solid line $(\alpha=0.002)$ and a dash line $(\alpha=0.1)$, respectively. It can be observed that when $\alpha$ is less than 0.002, the longitudinal velocity of 1-3, 1-3-2 and 2-1-3 composites almost remains unchanged. Within this range, the composites can be treated as fine-scaled. However, when $\alpha$ is larger than 0.002, which means the lateral scale is getting coarser and coarser, the longitudinal velocity of 1-3 and 1-3-2 composites will decrease quickly and reach their minima, at which $\alpha$ is about 2. While the longitudinal velocity of 2-1-3 composite will still remain the same until $\alpha$ gets to 0.1. In other words, for the same longitudinal velocity, 2-1-3 composite is of a lower fineness requirement of manufacturing compared with 1-3-2 composite.

\subsection{Comparison of thickness electromechanical coupling coefficient}
\begin{figure}[!b]
\centerline{\includegraphics[width=0.9\textwidth]{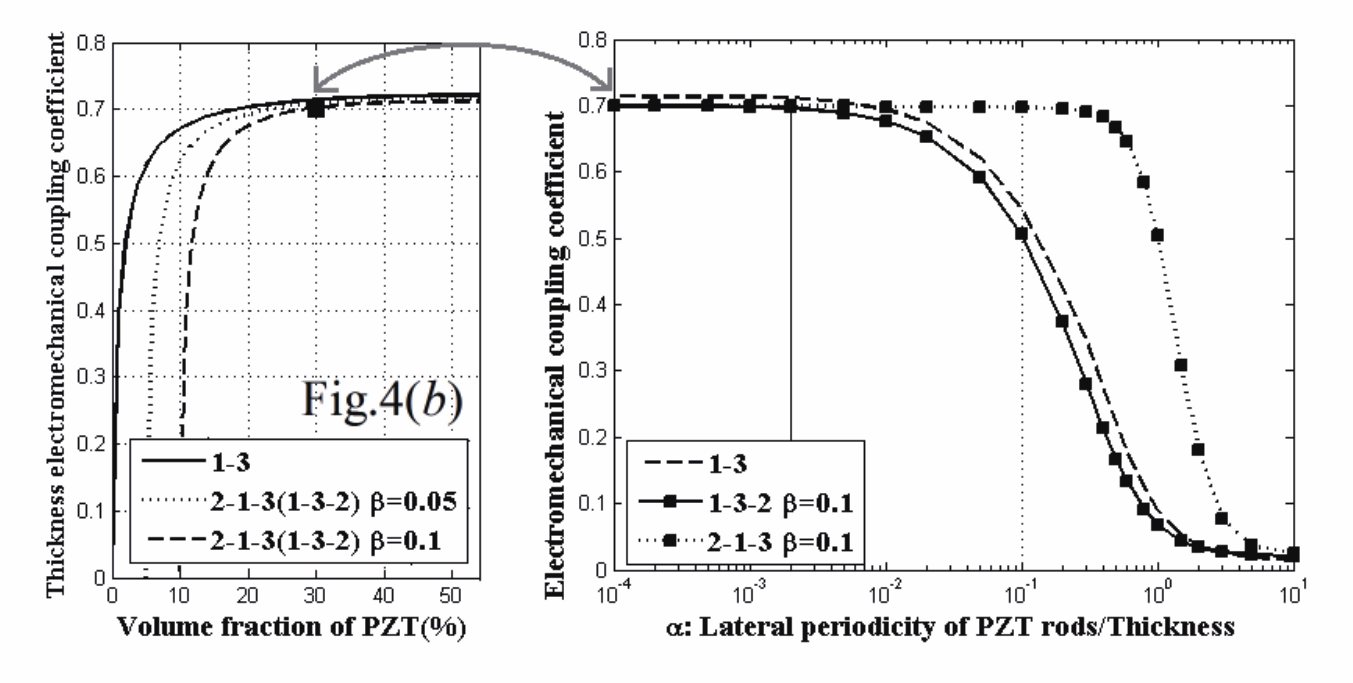}}
\caption{(Colour online) The dependence of thickness electromechanical coupling coefficient of 1-3, 1-3-2 and 2-1-3 composites $(VFP=30\%)$ on the ratio $\alpha$.} 
\label{figure-9}
\end{figure}

The dependence of the thickness electromechanical coupling coefficient $k_{\textrm{t}}$ on the ratio $\alpha$ is calculated and the result is shown in figure~\ref{figure-9}. Similarly, there are two solid squares corresponding to each other and representing $k_{\textrm{t}}$ of 30\%-VFP fine-scaled 1-3-2 or 2-1-3 composite $(\beta=0.1)$. The effect of the ratio $\alpha$ on the thickness electromechanical coupling coefficient is similar to the effect of $\alpha$ on the longitudinal velocity.

The thickness electromechanical coupling coefficient $k_{\textrm{t}}$ can reflect the efficiency of energy conversion of a transducer vibrating in a thickness mode. The higher  is $k_{\textrm{t}}$, the stronger  is the piezoelectricity. $k_{\textrm{t}}$ of 30\%-VFP fine-scaled 1-3, 1-3-2 and 2-1-3 composites are nearly the same (about 0.7) as shown in figure~\ref{figure-4}~(b). In order to make $k_{\textrm{t}}$ reach 0.7, the aspect ratio $\alpha$ of 1-3-2 composite should be no larger than 0.002, whereas $\alpha$ of 2-1-3 composite only need to be no larger than 0.1. That is to say, for 10~mm-thickness composites ($\beta=0.1$), the lateral periodicity of piezoelectric rods of 1-3-2 composite should be no larger than $20$~\textmu{}m while the lateral periodicity of 2-1-3 composite only needs to be no larger than 1~mm. The obtained largely reduced fineness requirement of manufacturing is an obvious advantage of 2-1-3 composite compared with 1-3 and 1-3-2 composites. From another perspective, when the aspect ratio $\alpha$ varies from 0.002 to 5, $k_{\textrm{t}}$ of 2-1-3 composite is obviously higher than that of 1-3 and 1-3-2 composites for a certain aspect ratio. For example, when the aspect ratio $\alpha$ is  equal to 1, $k_{\textrm{t}}$ of 2-1-3 composite is about 5.58 times that of 1-3 composite and 7.42 times that of 1-3-2 composite.

\subsection{Comparison of voltage receiving sensitivity}

\begin{figure}[!b]
	\centerline{\includegraphics[width=0.88\textwidth]{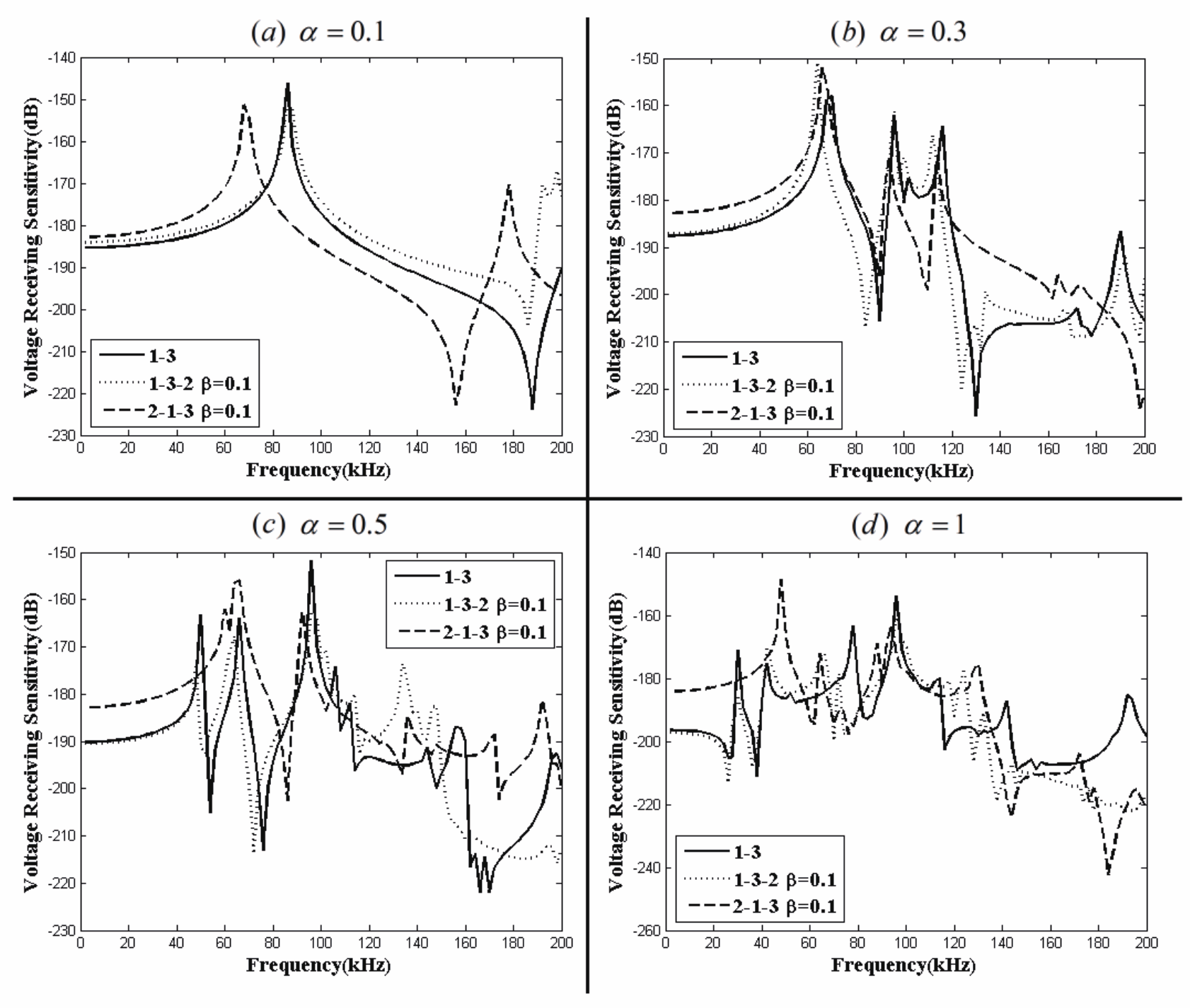}}
	\caption{The voltage receiving sensitivity (VRS) of 1-3, 1-3-2 and 
		2-1-3 composite transducers (VFP=30\%, $\beta$=0.1) (a) $\alpha$=0.1, (b) $\alpha$=0.3, (c) $\alpha$=0.5, (d) $\alpha$=1.} 
	\label{figure-10}
\end{figure}

In order to compare the voltage receiving sensitivity of transducers made of 1-3, 1-3-2 and 2-1-3 composites, the thickness of transducers is set as 10~mm and the ratio $\beta$ of 1-3-2 and 2-1-3 composites is set as 0.1. Several representative values of the ratio $\alpha$ $(0.1, 0.3, 0.5, 1)$ are chosen according to the results of longitudinal velocity $v_{\textrm{l}}$ and thickness electromechanical coupling coefficient $k_{\textrm{t}}$ shown above. When $\alpha$ is within the range from 0.1 to 1, there is a big difference between 2-1-3 composite and 1-3 or 1-3-2 composite in terms of $v_{\textrm{l}}$ and $k_{\textrm{t}}$. The results of voltage receiving sensitivity (VRS) from 0~Hz to 200~kHz are shown in figure \ref{figure-10}.

For a receiving transducer, the higher is the first-order resonant frequency (FRF), the wider is the frequency band having flat VRS. The higher is VRS, the better  is the performance of the receiving transducer. That is to say, it is hoped that the transducer is of high and stable VRS within a wide frequency band. 

From the results shown in figure \ref{figure-10}, it can be observed that for a certain $\alpha$, the results of 1-3 and 1-3-2 composites are similar while the result of 2-1-3 composite is quite different from those of the former two. The first-order resonant frequency (FRF) and VRS at 2~kHz in figure~\ref{figure-10} are listed in table~\ref{table-2}. 

It can be concluded that with an increase of $\alpha$, FRF of 1-3 and 1-3-2 composite transducers decrease a lot while a decrease of FRF of 2-1-3 composite transducer is much lower. Within the frequency band lower than FRF, VRS is almost flat. Therefore, for composites with large aspect ratio $\alpha$, 2-1-3 composite transducer is of a wider frequency band within which VRS is flat. With an increase of $\alpha$, VRS at 2~kHz of 1-3 and 1-3-2 composite transducers obviously decreases while that of 2-1-3 composite transducer almost keeps the same. The voltage receiving sensitivity at 2~kHz of 2-1-3 composite transducer is 13.1~dB higher than that of 1-3-2 composite transducer and 12.3~dB is higher than that of 1-3 composite transducer. Hence, 2-1-3 composite transducer shows a better performance when the lateral spatial scale of the composite should turn to coarse in order to loosen the fineness requirement of the manufacturing.

\begin{table}[!t] 
\footnotesize 
\caption{Resonant frequency and Voltage receiving sensitivity.}  
\begin{center}  
\renewcommand{\arraystretch}{1.25}	
\begin{tabular}{c|c|c|c|c|c|c}  
\hline \hline  
\multirow{3}{*}{$\alpha$} & \multicolumn{3}{|c|}{First-order resonant frequency (kHz)} & \multicolumn{3}{|c}{Voltage receiving sensitivity at 2~kHz (dB)}\\  
\cline{2-7}  
 & 1-3 composite & 1-3-2 composite & 2-1-3 composite & 1-3 composite & 1-3-2 composite & 2-1-3 composite\\
\hline
\cline{2-7}  
0.1 & 86 & 86 & 68 & $-$185.4 & $-$184.1 & 182.8 \\ 
\cline{1-7}  
0.3 & 70 & 64 & 66 & $-$187.6 & $-$187.1 & $-$182.8 \\ 		
\cline{1-7}  
0.5 & 50 & 48 & 60 & $-$190.2 & $-$190.6 & $-$182.9 \\ 		
\cline{1-7}  
1 & 30 & 30 & 48 & $-$196.5 & $-$197.3 & $-$184.2 \\ 	
\hline \hline  
\end{tabular}  
\end{center} 
\label{table-2}
\vspace{-3mm} 
\end{table}

\section{Root cause analysis}

\begin{figure}[!b]
\centerline{\includegraphics[width=0.8\textwidth]{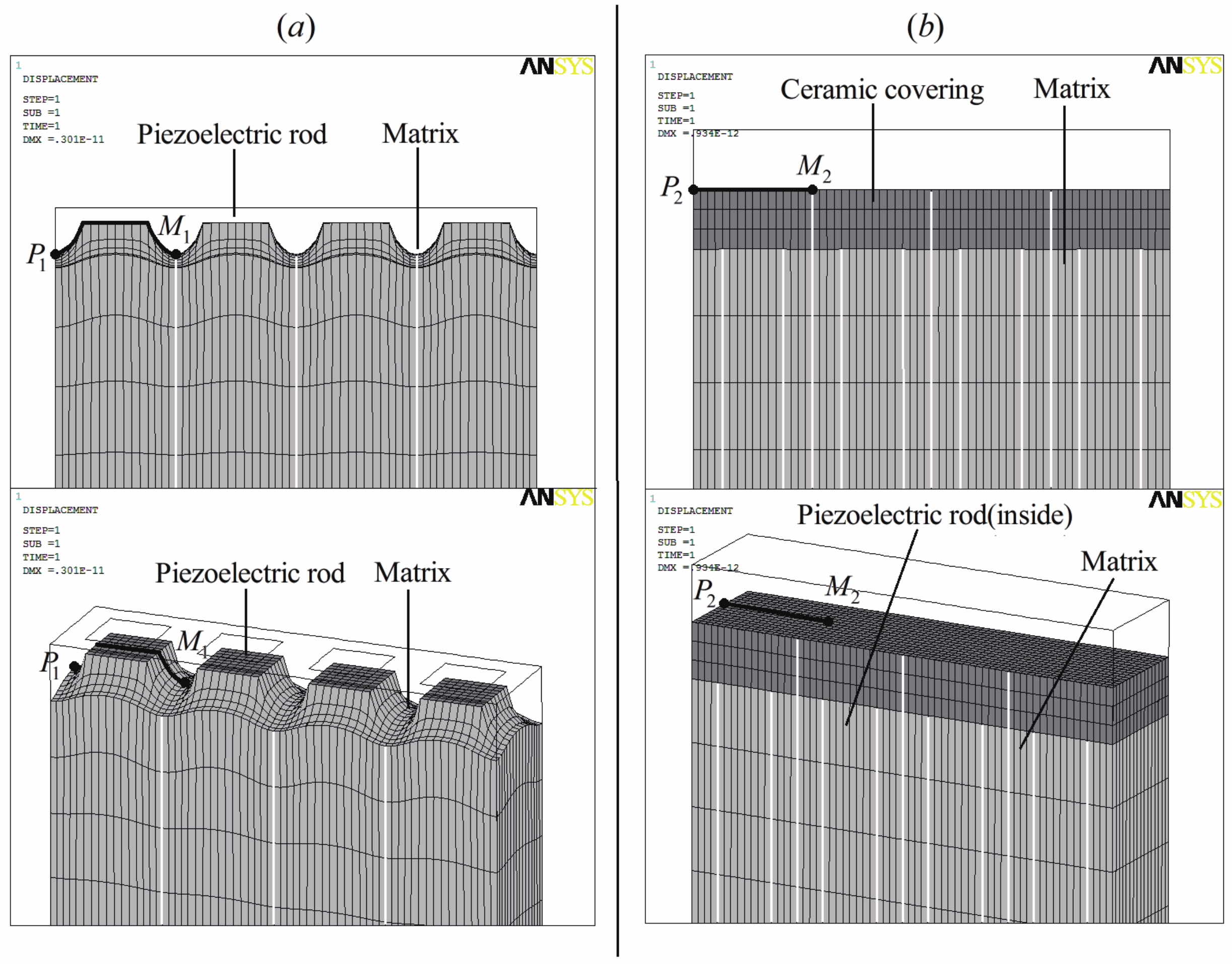}}
\caption{(Colour online) Mode animation figure of (a): 1-3-2 composite, (b): 2-1-3 composite 
(four unit cells combined, partially showed, 30\%-volume fraction, $\alpha$=0.1, $\beta$=0.05).} 
\label{figure-11}
\end{figure}

The main reason for the properties of coarse-scaled 2-1-3 composite being quite different from those of coarse-scaled 1-3-2 composite is the difference of their stress surfaces. The stress surface of 1-3-2 is the same as that of 1-3 composite. Both of them are a plane where piezoelectric phase and polymer phase are disjunctively distributed. It is obvious that a soft matrix will show a larger compression deformation than hard piezoelectric rods when the same pressure is applied to them. The effect of the soft matrix on the stiffness of the composite is stronger than that of hard piezoelectric rods. As a result, the longitudinal wave velocity of the composite which is determined by stiffness and density will be lower. Since much of the pressure applied to the composite is squandered compressing the passive phase instead of being applied effectively to the piezoelectric rods that are capable of conducting energy transformation, the piezoelectricity will become weaker.

The stress surface of 2-1-3 composite is the surface of a ceramic layer and is firmly supported by the piezoelectric rods under it. This structure is capable of applying more  force on the stress surface to the piezoelectric rods and reducing the waste in the polymer phase. The compression deformations both in piezoelectric phase and in matrix phase are approximately equal, namely the ``same strain in both phases'' condition of fine-scaled composite is approximately satisfied. Therefore, the property of coarse-scaled composite is nearly the same as that of a fine-scaled composite. The longitudinal velocity and the thickness electromechanical coupling coefficient are larger and the voltage receiving sensitivity also shows a better performance. Hence, the fineness requirement of manufacturing can be reduced without losing good piezoelectricity and sensitivity. 

\begin{figure}[!t]
\centerline{\includegraphics[width=0.62\textwidth]{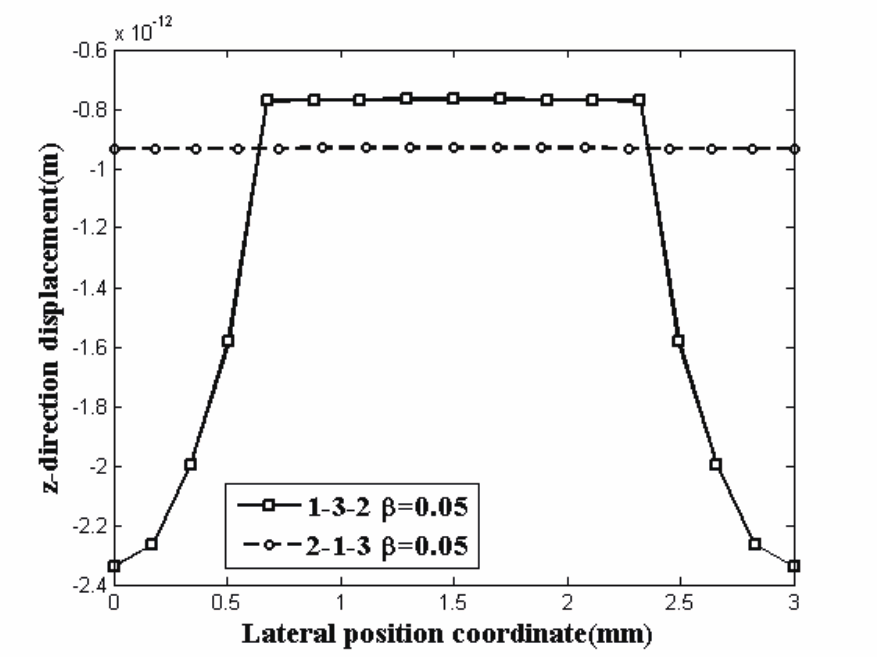}}
\caption{$z$-direction displacement distribution on line $P_1M_1$ of 1-3-2 composite 
and $P_2M_2$ of 2-1-3 composite in figure \ref{figure-11}.} 
\label{figure-12}
\end{figure}

The strain distributions on surfaces of 1-3-2 and 2-1-3 composites under the pressure load will be shown to verify the above statement. Four unit cells of the composite are combined and presented. 30\%-volume fraction is chosen and the ratio $\alpha$ is set as 0.1 and $\beta$ is set as 0.05. After static analysis, mode animation is taken to observe the strains of both phases as shown in figure~\ref{figure-11}~(a) and (b). Figure~\ref{figure-12} depicts $z$-direction displacement distribution on line $P_1M_1$ of 1-3-2 composite and $P_2M_2$ of 2-1-3 composite. As seen in figure~\ref{figure-11}~(a) and figure~\ref{figure-12}, on the stress surface of 1-3-2 composite, the $z$-direction displacement of matrix phase is obviously larger than that of piezoelectric phase. Figure~\ref{figure-11}~(b) and figure~\ref{figure-12} show that the $z$-direction displacement of the ceramic covering layer is almost the same all over the stress surface.

\section{Conclusions}
In this paper, a systematically comparative study on 1-3, 1-3-2 and 2-1-3 piezoelectric composite transducers is presented. The finite element method (FEM) has been effectively used to model the composites and calculate their performance parameters including longitudinal velocity, thickness electromechanical coupling coefficient and voltage receiving sensitivity. The particular advantages of 2-1-3 composite serving as a kind of receiving transducer material are highlighted.

The composites are divided into two major categories, fine-scaled and coarse-scaled, according to the lateral spatial scale of piezoelectric rods or the aspect ratio $\alpha$. Fine-scaled composites meet the ``same strain in both phases'' condition while coarse-scaled composites do not. This leads to the different research methods of these two categories, namely theoretical derivation and FEM, respectively. 

The longitudinal velocity and the thickness electromechanical coupling coefficient of fine-scaled composite transducers are determined by the properties and volume fractions of two phases and have nothing to do with the aspect ratio. It is concluded that fine-scaled 1-3-2 composite and fine-scaled 2-1-3 composite are of the same electro-elastic properties, which are similar to those of fine-scaled 1-3 composite.

The advantages of 2-1-3 piezoelectric composite show up in coarse-scaled composite group. In terms of longitudinal velocity and thickness electromechanical coupling coefficient, when the aspect ratio $\alpha$ varies from 0.002 to 5, the performance of 2-1-3 composite is better than that of 1-3 and 1-3-2 composites for a certain aspect ratio. The thickness electromechanical coupling coefficient of 2-1-3 composite is about 5.58 times that of 1-3 composite and 7.42 times that of 1-3-2 composite. From a different perspective, when the three are of the same performance, 2-1-3 composite is of a larger aspect ratio and, therefore, of a lower manufacturing difficulty. In the aspect of voltage receiving sensitivity (VRS), the performances of 1-3 and 1-3-2 composite transducers are similar while that of 2-1-3 composite transducer is significantly better. For a large aspect ratio $\alpha=1$, 2-1-3 composite transducer is of a wider flat-VRS frequency band. At the frequency 2~kHz, the VRS of 2-1-3 composite transducer is 13.1~dB higher than that of 1-3-2 composite transducer and 12.3~dB higher than that of 1-3 composite transducer.

The advantages of 2-1-3 piezoelectric composite transducer are due to its particular transverse and longitudinal piezoelectric support structure, which is capable of effectively carrying out electro-mechanical transformation.

%
%

\ukrainianpart
\title{Порівняння ефективності композитних п'єзоелектричних перетворювачів типу 2-1-3, 1-3 і 1-3-2 
	   методом скінченних елементів}
\author{Я. Сан\refaddr{label1}, Б. Хуа\refaddr{label2}}
\addresses{
	\addr{label1} Коледж природничих наук, Пекінський лісотехнічний університет, Пекін, 100083, Китай
	\addr{label2} Науково-дослідний інститут інженерних систем, Китайська державна суднобудівна корпорація, \\ Пекін, 100094, Китай}

\makeukrtitle

\begin{abstract}
	П'єзоелектричні композити типу 1-3, 1-3-2 і 2-1-3  --- це три відповідні розумні матеріали для конструювання та виробництва ультразвукових 
	перетворювачів. Вони запропоновані для використання в різних каскадах, але володіють подібними властивостями.  Композит 1-3-2 володіє вищою
	механічною стійкістю у порівнянні  з композитом типу  1-3.  Композит 2-1-3 легший у виробництві у порівнянні з композитом  1-3-2. 
	У цій статті представлено порівняльне дослідження  цих трьох композитів з огляду на властивості матеріалу приймального перетворювача.
	Метод скінченних елементів (МСЕ) застосовано для обчислення поздовжньої швидкості,  товщинного коефіцієнта  електромеханічного зв'язку  та 
	напругової чутливості прийому. Зроблено висновок, що при великому відношенні сторін  $\alpha=1$, ефективність композитного перетворювача  
	значно краща, ніж характеристики композитних  перетворювачів  1-3 і 1-3-2. Товщинний коефіцієнт електромеханічного зв'язку композитного
	перетворювача 2-1-3 приблизно  у  5.58 разів  більший за коефіцієнт композитного перетворювача  1-3 
	та у 7.42 разів більший за коефіцієнт  композитного перетворювача  1-3-2. 
	Напругова чутливість прийому  при  2~кГц композитного перетворювача 2-1-3 на 13.1~дБ вища за напругову чутливість композитного перетворювача
	1-3-2 та на  12.3~дБ вища за напругову чутливість композитного перетворювача  1-3.
	\keywords п'єзоелектричні матеріали, п'єзоелектричний перетворювач, скінченний елемент,  PZT кераміка, композити на полімерній основі, 
	п'єзоелектрика
\end{abstract}

\end{document}